\documentclass[a4paper,11pt]{article}
\usepackage[a4paper,top=2cm,bottom=2cm,left=3cm,right=3cm,marginparwidth=1.75cm]{geometry}
\usepackage[pdftex, pdftitle={Article}, pdfauthor={Author}]{hyperref}
\usepackage[utf8]{inputenc}
\usepackage{float,array,multirow}
\usepackage{subcaption}
\usepackage{overpic}   
\usepackage{todonotes}
\usepackage{amsmath}
\usepackage{overpic}
\usepackage{amssymb}
\usepackage[english]{babel}
\usepackage[dvipsnames]{xcolor}
\usepackage{pifont}
\usepackage{graphicx}
\usepackage{wrapfig}

\usepackage{placeins}
\usepackage{lscape}
\usepackage{rotfloat}

\usepackage[backend=biber,maxnames=2,style=numeric-comp,sorting=none]{biblatex}
\DeclareFieldFormat[article]{pages}{#1}
\DeclareBibliographyAlias{article}{std}
\DeclareBibliographyAlias{online}{std}
\DeclareBibliographyAlias{book}{std}
\DeclareBibliographyDriver{std}{
  \usebibmacro{bibindex}
  \usebibmacro{begentry}
  \usebibmacro{author/editor+others/translator+others}
  \setunit{\labelnamepunct}\newblock
  \newunit\newblock
  \usebibmacro{date}
  \newunit\newblock
  \usebibmacro{journal}
  \newunit\newblock
  \printfield{volume}
  \newunit\newblock
  \printfield{pages}
  \newunit\newblock
  \usebibmacro{finentry}}

\begin{document}
\clearpage\thispagestyle{empty}
\vspace{3cm}
\begin{center}
\section*{Assessing the effect of error field penetration during plasma current ramp-up in the DIII-D tokamak}
\vspace{0.8cm}
C. F. B. Zimmermann\textsuperscript{1 *}, 
E. M. Bursch\textsuperscript{1}, 
C. Paz-Soldan\textsuperscript{1},\\
J. M. Hanson\textsuperscript{1}, 
N. Leuthold\textsuperscript{1},
N. C. Logan\textsuperscript{1},
A. O. Nelson\textsuperscript{1}\\
\vspace{0.6cm}

\textit{\textsuperscript{1} Columbia University, New York, USA}\\

\vspace{0.6cm}
\textsuperscript{*} E-mail of the corresponding author: benedikt.zimmermann@columbia.edu\\
\end{center}

\begin{abstract}
\noindent This work provides evidence that established error field penetration threshold scalings remain applicable during plasma current ramp-up. In dedicated DIII-D experiments with imposed $n=1$ perturbations during extended $I_p$ ramps, an apparent empirical threshold is found between $2$ and $3$~kA of applied 3D coil current, above which MHD modes are seeded. The imposed perturbation couples to the rational surfaces present during the ramp, seeding near the $q=4$ surface and penetrating as an $m/n=3/1$ mode by the end of the perturbation phase. To interpret these observations, multi-machine penetration threshold scalings are combined with equilibrium-based overlap metrics from the GPEC code, including the in-situ error fields of the device. This modeling reproduces the observed onset in the amplitude scan and classifies mode seeding across a database of 12 ramp-up discharges spanning a range of plasma currents and densities. Across this database, the seeding seems to be controlled by the applied 3D coil current rather than by the plasma current or its ramp rate. Accounting for the in-situ error fields is found to be important for reliable prediction. These results are consistent with the robustness of scaling-based penetration metrics when coupled to detailed 3D field modeling under transient ramp-up conditions, and suggest the importance of accounting for in-situ error fields when assessing additional externally induced perturbations. This work is motivated by future tokamaks in which transient, non-axisymmetric error fields can arise during startup, for example from closed-loop, passive runaway electron mitigation coils energized by loop voltage ramping up plasma current.
\end{abstract}

\section{Introduction}

Disruption mitigation schemes will be essential for future high-current tokamaks, as disruptions impose significant thermal and mechanical loads on tokamak components \cite{Loarte_2011,Lehnen_2015,Breizman_2019,Sweeney_2020,Ratynskaia_2026,Sweeney_2026}. A promising way to mitigate the resulting runaway electron (RE) beams is to surround the plasma vessel with closed-loop, passively conducting structures. During a current quench, strong currents are induced in such a runaway electron mitigation coil (REMC), producing a large non-axisymmetric perturbation of the magnetic field topology that ultimately deconfines the REs \cite{Boozer_2011,Smith_2013_REMC,Breizman_2019,Weisberg_2021,Tinguely_2021,Battey_2024,Levesque_2026}.\par
The REMC is one example of a broader effect: any change in plasma current induces currents in the surrounding conducting structures, and these in turn perturb the magnetic field topology. In routine operation such currents arise in the vacuum vessel and in-vessel structures and from coil misalignments or ferromagnetic components, producing non-axisymmetric error fields (EFs). These contributions are especially consequential early in the ramp-up, when current transients are strongest. Even small in-situ EFs of order $\delta B/B\approx10^{-4}$ can penetrate the magnetic topology. Error field \textit{penetration}, the transition from a small, largely screened non-axisymmetric perturbation to a large magnetic island, is therefore a critical issue for both present-day and future devices \cite{Hender1992,Buttery1999,Scoville2003,Wolfe2005,Park2007,Howell2007,Menard2010,Park_2011}. In more detail, penetration occurs as the plasma is slowed at surfaces resonant with the perturbation: depending on the plasma scenario \cite{Fitzpatrick1991,Fitzpatrick1993,Cole2006,logan2020robustness,Peterka_2024}, there is a threshold external error field above which the plasma rotation can no longer counteract the electromagnetic torque, so the plasma comes to rest and magnetic islands form, ultimately leading to harmful disruptions.\par
Recognized early as a significant risk to future reactor scenarios, this process has motivated substantial effort to study the penetration threshold and the in-situ error fields of present-day devices \cite{LaHaye1992,LaHaye1991_RSI,Garofalo2002,Luxon2003,Park_2011,Wang2016,Wolfe2005_CMod,Buttery2000,Menard2010_NSTX,Maraschek2013,Bandyopadhyay_2025}. The penetration-threshold scalings developed from these studies (detailed in Section \ref{sec:methods}) have been validated primarily for the plasma flat top, where they describe the onset of low toroidal mode numbers and their coupling to the correspondingly lowest $q$ (safety factor) surfaces \cite{logan2020robustness}. During a transient ramp-up, however, the skin effect causes the induced electric field to drive current predominantly at the plasma edge, flattening the current profile so that higher-$q$ surfaces enter the plasma center early. In principle, the coupled evolution of the current profile, plasma rotation, and non-axisymmetric plasma response during such a transient can be captured self-consistently by time-dependent nonlinear MHD codes \cite{Jardin_2007,Sovinec_2004,Czarny_2008,Hoelzl_2021,Yu_2003}. Such simulations are, however, computationally demanding, which makes them ill-suited to rapidly scoping the many candidate scenarios encountered in the design and operation of future devices. This work therefore takes a deliberately lighter approach and investigates whether the far cheaper, quasi-steady penetration-threshold scalings, combined with equilibrium-based 3D-field modeling, remain applicable under transient current ramps dominated by higher-$q$ surfaces, thereby providing a practical means to scope error-field penetration limits without a full time-dependent treatment. We find that they do: in a dedicated amplitude scan, an apparent empirical penetration threshold appears between $2$ and $3$~kA of applied $n=1$ field, and the combined scaling+GPEC workflow reproduces this onset and classifies mode activity across a database of ramp-up discharges.\par
The paper is structured as follows. Section \ref{sec:methods} describes the penetration-threshold framework, the scaling laws and 3D-field modeling, and the magnetic diagnostics used in this work. Section \ref{sec:analysis} presents a detailed experimental analysis and modeling of a selected set of discharges with varying imposed 3D fields during the ramp-up. Section \ref{sec:database} extends the analysis to a larger database of discharges to strengthen confidence in the results. Section \ref{sec:summary} concludes.

\section{Methodology}\label{sec:methods}
The typical experimental approach to characterizing penetration applies a controllable non-axisymmetric field to a stable background scenario and studies the emergence of modes, e.g., by measuring the response of the magnetic topology. The threshold is then usually defined experimentally by the magnitude of the current generating the perturbation and the phase of this perturbation in the rest frame of the device.\par
To design stable operating points for future devices, \textit{error field penetration threshold scalings} have been derived from multi-machine databases of such experiments \cite{logan2020robustness,logan2020empirical,bursch_2026}. The threshold is typically expressed via the \textit{dominant mode overlap metric} $\delta$, which quantifies the non-axisymmetric field relative to the axisymmetric field \cite{Park_2011,Buttery_2012,Pharr_2026}. Specifically, $\delta$ is the projection of the area-normalized external field spectrum onto the dominant coupling mode of the plasma, normalized by the toroidal field $B_t$. Extensive work has related observed thresholds in the overlap metric to background plasma parameters, e.g. in Ref.~\cite{logan2020robustness}. In this work, we use the more recent formulation of Ref.~\cite{bursch_2026}, obtained from the multi-machine ITPA error field penetration database, which was extended to include data from Alcator C-mod, DIII-D, EAST, JET, J-TEXT, and KSTAR:
\begin{equation}
    \delta_\text{thresh.} = 10^{-4.26\pm0.09}\left(\frac{\beta_n}{l_i}\right)^{0.13\pm 0.06}|I_p|^{-1.01\pm0.07}R_0^{1.57\pm 0.15} n_e^{0.56\pm 0.08} |B_t|^{0.30\pm 0.10} \text{.}
    \label{equ:Bursch_scaling}
\end{equation}
Here, $n_e$ is the line-averaged electron density in $10^{19}$ m${}^{-3}$, $B_t$ the toroidal magnetic field in T, $R_0$ the major radius in m, $I_p$ the plasma current in MA, $\beta_n$ the normalized plasma pressure, and $l_i$ the internal inductance. These laws show, for example, that for a single device (fixed $B_t$ and $R_0$) density and current are the dominant control parameters.\par
A key step is to compare the predicted threshold $\delta$ (from the scaling) with the imposed $\delta$, e.g. from externally applied non-axisymmetric coil fields or the device's in-situ error fields. In practice, the imposed $\delta$ is computed from the background plasma conditions, equilibrium reconstruction, the applied 3D coil configuration, and machine-specific error field models using the GPEC code \cite{Park2007,Park2009_Shielding,Park_2011}. For the DIII-D tokamak, the SURFMN code~\cite{Schaffer2008_ITERcoils} models the contribution of in-situ error fields due to the measured misalignment of coils and the toroidal field coil current feeds, whereas induced currents in passive structures are not included, as this would require more sophisticated eddy current modeling. It should be emphasized, however, that these scalings carry substantial uncertainties and capture trends rather than a precise threshold, as reflected in the sizable exponent uncertainties in Eq.~\ref{equ:Bursch_scaling}. The scaling is therefore best interpreted probabilistically, defining a range of imposed $\delta$ over which locking becomes increasingly likely \cite{Pharr_2024,Logan_2026}. When the imposed $\delta$ exceeds the threshold, EF penetration is expected.\par
Experimentally, the emergence or seeding of a mode is detected with 3D magnetic sensors around the device that resolve changes in $B_\theta$ and $B_r$. A Fourier decomposition of these measurements extracts the toroidal and poloidal mode components of the perturbations, with $n$ and $m$ denoting the toroidal and poloidal periodicity. Careful subtraction and compensation schemes are required to account for finite pick-up of background fields and of the applied perturbations. Further details on the DIII-D magnetic diagnostics and the corresponding analysis techniques are given in Ref. \cite{Strait_2006,King_2014,Strait_2016_RSI}.\par

\section{Detailed Analysis}
\label{sec:analysis}

\subsection{Experimental Analysis}
For this work, DIII-D discharges $170{,}000-200{,}000$ were scanned to identify experiments that simultaneously exhibited significant currents above $2$~kA in the in-vessel (I-coils) or ex-vessel (C-coils) 3D coils \cite{Scoville_2003}, strong $\dot{I}_p$ indicative of current ramps, and 3D coils operating out of phase with the typical error field correction. This yielded a relatively small set of discharges suitable for detailed study. In particular, a subset of four discharges was identified in which rectangular I-coil waveforms of increasing amplitude were applied during an extended current ramp-up. Here the C-coils provided error field correction, so the 3D field from the I-coils is the dominant error field acting on the plasma.

At the chosen sampling time ($0.9$~s), the four discharges have the following parameters: line-averaged density $n_e = 1.86\times10^{19}\,\mathrm{m^{-3}}$, toroidal field $B_t = 2.15\,\mathrm{T}$, major radius $R_0 = 1.72\,\mathrm{m}$, normalized beta $\beta_n = 0.25$, internal inductance $l_i = 0.90$, minor radius $a = 0.57\,\mathrm{m}$, plasma current $I_p = 1.07\,\mathrm{MA}$, elongation $\kappa = 1.91$, and current ramp rate $\dot I_p = 0.87\,\mathrm{MA/s}$. As these are repeat discharges, the values are overall reproducible across the set. The background plasma and equilibrium parameters differ only minimally between the four discharges, so the deliberately varied I-coil current should dominate any mode seeding and onset.

\begin{figure}
    \centering
    \includegraphics[width=0.7\linewidth]{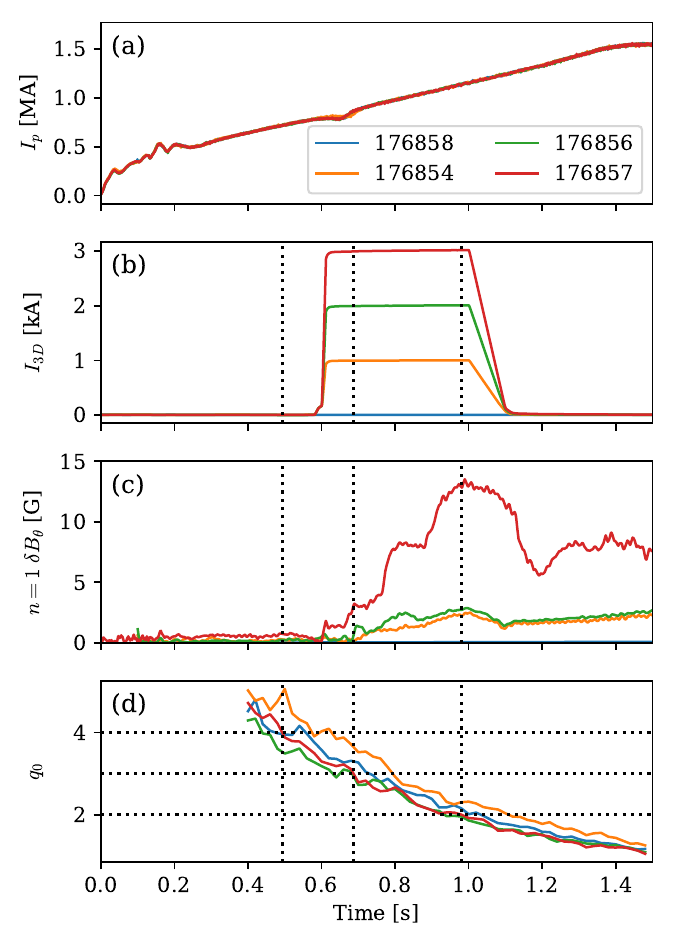}
    \caption{Key time traces for the selected discharges: (a) plasma current $I_p$, showing the extended ramp-up; (b) applied $n=1$ I-coil currents with amplitudes of $0$, $1$, $2$, and $3$~kA; (c) perturbed poloidal field $\delta B_\theta$ reconstructed from magnetic measurements; and (d) on-axis safety factor from EFIT constrained by MSE measurements. Vertical dotted lines mark the approximate times at which $q_0=4,3,2$ is reached, i.e. when the corresponding rational surfaces enter the plasma.}
    \label{fig:timetraces}
\end{figure}

The most important time traces for this subset are shown in Fig.~\ref{fig:timetraces}. Panel (a) shows an extended current ramp-up to $I_p = 1.55$~MA over roughly $1.5$~s, corresponding to $\dot{I}_p \approx 1$~MA/s. Panel (b) shows the I-coil waveforms, applied for about $0.4$~s between $0.6$ and $1.0$~s, with currents of $0$, $1$, $2$, and $3$~kA across the four discharges. The displayed current represents the $n=1$ component of the perturbation. Higher toroidal mode numbers were minimized, giving very low $n=2$ and $n=3$ amplitudes of the imposed 3D field.

Before the mode traces in Panel (c) can be interpreted, the analysis must isolate the plasma response from background pickup. For this study, an analysis scheme based on pair-balanced differential probes removed the $n=0$ component from the magnetic mode decomposition, and background subtraction used the final $0.1$~s before the current ramp (i.e. $-0.1$–$0$~s). The mode decomposition was first applied to discharge \#176858, which has zero I-coil current. The decomposition is performed exclusively in the toroidal direction, i.e. the signals of toroidal probe arrays at fixed poloidal positions are fitted with $n=1,2,3$ harmonics in the geometric toroidal angle. This yielded slowly drifting $n=1,2,3$ components of the perturbed poloidal field $\delta B_\theta$, attributed to residual pickup from the axisymmetric $I_p$ and to integrator drift \cite{Strait_2006} rather than MHD activity. This discharge was therefore used as a reference and subtracted from the magnetic signals of the other discharges prior to analysis. Empirical models of the I- and C-coils from vacuum discharges were additionally used to subtract their direct coupling to the magnetic measurements. However, coupling between the I-coils and the midplane $B_\theta$ sensors is generally weak due to the coil geometry. Together, these steps isolate the 3D plasma response in the transient scenario studied. As an independent cross-check, the analysis was repeated with the magnetics analysis modules of the OMFIT framework~\cite{Meneghini_2015}, which implement a similar toroidal decomposition with pickup compensation, yielding very similar results for \#186857.

The measured $n=1$ $\delta B_\theta$ mode activity, Panel (c), is shown for all four discharges at the high-field-side midplane. The high-current case ($\approx 3$~kA, red) shows a significant mode amplitude that grows as the 3D coils are ramped up and persists after they are ramped down. The low- and medium-current cases (orange and green) show only weak signals during the perturbation and a small ramp-like signature afterward, both attributed to residual pickup and noise not fully removed by the analysis. Dedicated vacuum discharges with the specific I-coil waveforms of this experiment could further reduce these contributions in future work, while part of the remaining signal may reflect the linear plasma response to the applied field not resulting in mode onset. Since neither the $1$~kA nor the $2$~kA case shows mode onset, whereas the $3$~kA case does, these observations suggest an empirical threshold in the externally imposed error field between $2$ and $3$~kA, above which mode onset occurs.

The on-axis safety factor $q_0$, Panel (d), is reconstructed by EFIT with current-profile constraints from motional-Stark-effect (MSE) measurements. In all discharges studied here, the on-axis value $q_0$ represents the minimum of the $q$-profile. For the MSE analysis, no $E_r$ correction was feasible due to unfavorable beam settings, but these discharges remain in L-mode confinement with toroidal rotation below $100$~km/s, so the MSE results are not strongly affected by rotation- or pressure-driven $E_r$ variations \cite{Burrell1997}. The reconstructed $q_0$ values are subject to non-negligible uncertainties, which can account for the apparent differences between the discharges, e.g. for \#176854 around $0.9$~s. A Monte Carlo analysis, in which the input data of the equilibrium reconstruction were varied within realistic experimental uncertainties, shows that the on-axis values remain consistent within these uncertainties. This is supported by the measured MSE pitch angles of the core channels, which are indistinguishable within the experimental noise for \#176856 and \#176857, i.e. the two discharges between which the threshold is observed. 

In the medium- and high-current cases, the $q=4$ surface enters the plasma before the perturbation, while the $q=3$ and $q=2$ surfaces enter during it. The $n=1$ component in the high-current case responds somewhat to the onset of the perturbation, so the mode could be seeded as an $m/n=4/1$ structure coupling to $q=4$. It then develops strongly once $q=3$ enters the plasma and likely evolves into a mixture of $m=4,3,2$ toward the end of the perturbation, coupling to lower rational $q$.\par
\begin{figure}
    \centering
    \begin{overpic}[width=0.7\linewidth,unit=1mm,grid=false]%
        {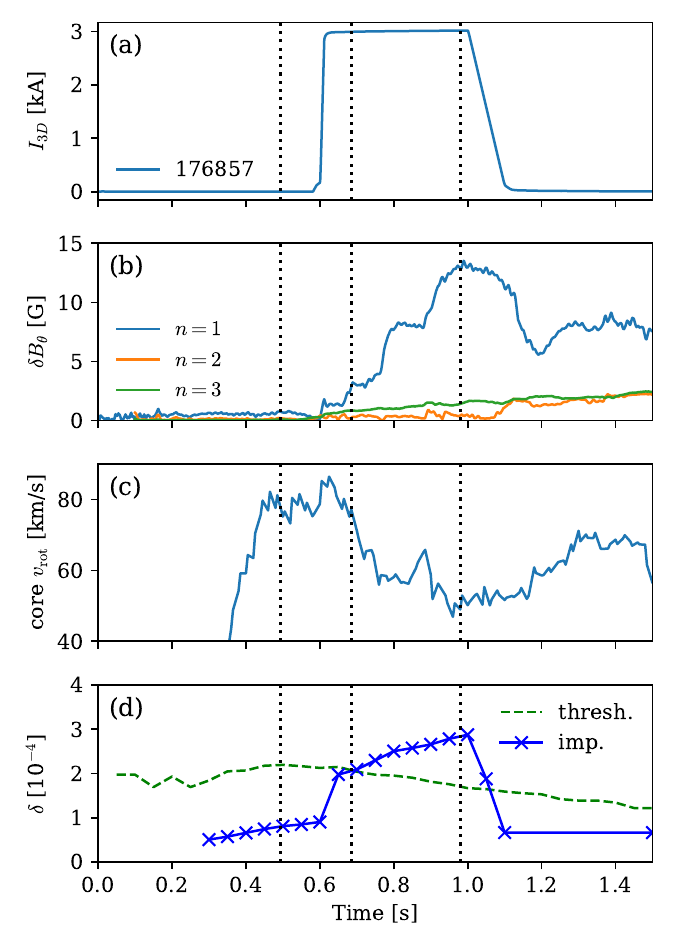}
        \put(26,99){$q_0{=}4$}
        \put(33.75,99){$q_0{=}3$}
        \put(45,99){$q_0{=}2$}
    \end{overpic}
    \caption{Toroidal mode number decomposition for the high-current $3$~kA discharge \#176857: (a) applied 3D perturbation waveform; (b) $n$-number decomposition of $\delta B_\theta$; (c) near-axis toroidal rotation; (d) time-dependent imposed (blue, with markers) and threshold overlap $\delta$. Vertical dotted lines mark the approximate times at which $q_0=4,3,2$ is reached in the high-current discharge \#176857, i.e. when the corresponding rational surfaces enter the plasma.}
    \label{fig:higher_n_numbers}
\end{figure}

The magnetic analysis was extended to higher toroidal mode numbers for all discharges, but only the high-current case yielded significant amplitudes. The full decomposition for discharge \#176857 is shown in Fig.~\ref{fig:higher_n_numbers}. That only the $n=1$ component grows, while higher-$n$ components stay near the noise floor throughout, confirms that the plasma responds selectively to the $n=1$ drive. Crucially, the $n=1$ amplitude rises mainly after $q_0=3$ enters the plasma, indicating that the $m/n=3/1$ resonant surface influences the onset of significant mode activity. The near-axis toroidal rotation in Panel~(c) shows this more explicitly, dipping once $q_0=3$ enters, in parallel with the growth of the $n=1$ mode. This correlated braking is the signature of an electromagnetic torque exerted on the plasma once the resonant mode is destabilized, slowing the core rotation as momentum is transferred to the mode. The imposed and threshold overlap parameter $\delta$ in Panel~(d) is discussed later in this section.

\begin{figure}[tbp]
    \centering
    \includegraphics[trim=2.5cm 0.5cm 2.5cm 0.5cm, width=0.4\linewidth]{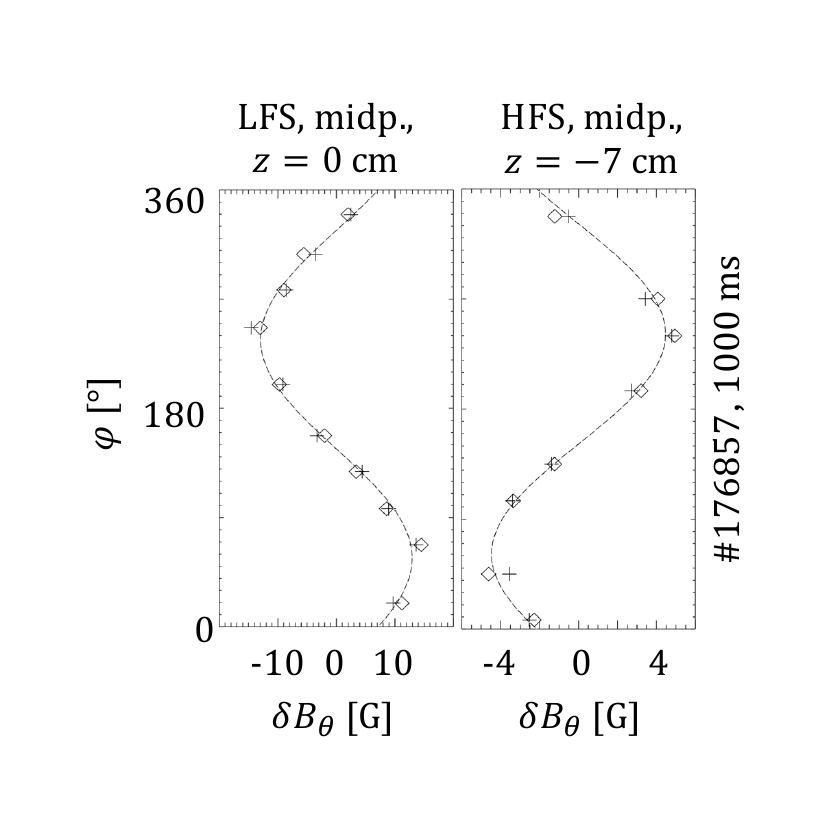}
    \caption{Toroidal variation of the perturbed poloidal magnetic field measured by probe arrays on the high-field-side (HFS) and low-field-side (LFS) midplane for the high-current discharge \#176857. Markers denote probe measurements and dashed lines the $n=1$ fits, where diamonds and $+$ symbols correspond to different sensors combined in a mutual correction scheme. Measured at the end of the perturbation at $1.0$~s, the HFS and LFS signals exhibit a $180^\circ$ phase shift, indicative of an odd $m$ number.}
    \label{fig:odd_vs_even}
\end{figure}

Finally, the poloidal mode numbers were investigated. Using the subtraction schemes explained above, the toroidal probe arrays on the high-field side (HFS) and low-field side (LFS) were analyzed, as shown for the high-current discharge \#176857 in Fig.~\ref{fig:odd_vs_even}. The HFS array sits at a vertical height of $z=-7$~cm and the LFS array at $z=0$~cm. The perturbed $\delta B_\theta$ is plotted against the toroidal angle. A clear $n=1$ toroidal periodicity is visible in all probe measurements (markers) and their fits (dashed lines). The plots show the measurement at $1.0$~s, just before the I-coils are switched off. The $180^\circ$ phase shift between HFS and LFS is clearly visible, indicative of an odd $m$ number and suggesting that at the end of the perturbation the mode exists on the $q=3$ surface as $m/n=3/1$.

The full 3D magnetic array was then used to estimate the poloidal mode number amplitudes quantitatively. Such a decomposition is not straightforward, as the increasing influence of wall and eddy currents must be accounted for in a reliable $m$-number analysis \cite{Schittenhelm_1997}. We therefore refrain from reporting quantitative values, but note that toward the end of the perturbation a mixture of $m=4,3,2$ is observed. More detailed analysis and modeling of this process is left for future work.

\subsection{Threshold Predictions}

To interpret these analysis results, the corresponding error field penetration threshold scalings are compared with stability calculations based on experimental equilibria. The comparison is shown in Panel~(d) of Fig.~\ref{fig:higher_n_numbers} for the high-current discharge \#176857. Two curves are plotted versus time: the threshold overlap $\delta$ predicted by Eq.~\ref{equ:Bursch_scaling} (green, dashed), and the imposed overlap $\delta$ (blue, with markers) computed with the GPEC workflow from the reconstructed equilibrium and the applied $n=1$ 3D coil field. Following the definition in Section~\ref{sec:methods}, the imposed $\delta$ is normalized by the toroidal field $B_t$ and therefore increases with the magnitude of the applied perturbation. Both evolve during the ramp-up as the background plasma parameters and the coil current change. Early in the perturbation the imposed $\delta$ lies below the predicted threshold, consistent with the absence of strong mode activity in Panels~(b) and (c). As the ramp proceeds, the imposed $\delta$ rises while the predicted threshold falls, and the two curves cross shortly after the $q=3$ surface enters the plasma. This crossing coincides in time with the growth of the $n=1$ mode and the rotation braking in Panels~(b) and (c), so the scaling-based workflow reproduces the timing and amplitude of the empirically observed onset. Importantly, the two ingredients carry independent information. The scaling of Eq.~\ref{equ:Bursch_scaling} supplies only the threshold magnitude, set mainly by $I_p$ and $n_e$, with only weak current-profile dependence through $(\beta_n/l_i)^{0.13}$. Moreover, the overlap metric $\delta$ has no explicit knowledge about the present rational surfaces and, therefore, quantifies a global drive of the plasma rather than the drive on an individual rational surface. The identification of the responding surface, $m/n=3/1$, comes instead from the magnetic measurements.\par

\begin{figure}
    \centering
    \includegraphics[width=0.5\linewidth]{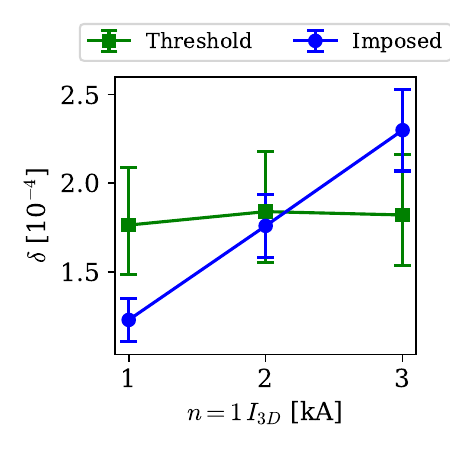}
    \caption{Comparison of the predicted error field penetration threshold and the experimentally calculated overlap metric $\delta$ as a function of applied 3D current. Green square markers show the threshold scaling from Eq.~\ref{equ:Bursch_scaling}, with uncertainty estimated via a Monte-Carlo variation of the scaling coefficients. Blue round markers show the overlap metric from GPEC calculations using experimental equilibria. The two curves intersect between $2$ and $3$~kA, indicating the predicted onset of mode penetration, consistent with the experimental observation at $3$~kA.}
    \label{fig:subset_scan_modeling}
\end{figure}

Expanding this analysis to the entire 3D coil current scan, Fig.~\ref{fig:subset_scan_modeling} shows the overlap metric $\delta$ as a function of 3D coil current amplitude, with data sampled again at $0.9$~s, while the $q=4$ and $q=3$ surfaces are in the plasma.

The predicted threshold (green curve, square markers, from Ref.~\cite{bursch_2026}, Eq.~\ref{equ:Bursch_scaling}) is nearly constant over the scan, as these are repeated discharges. This is also consistent with the stable time trace in Fig.~\ref{fig:higher_n_numbers}(d), which shows that the I-coil perturbation does not significantly modify the background plasma. As noted in the introduction, these threshold scalings usually carry large uncertainties, and two sources arise in the results presented here. The first is inherent to the scaling law itself, shown as error bars on the markers: a Monte Carlo approach evaluated the threshold overlap by varying the scaling coefficients of Eq.~\ref{equ:Bursch_scaling}, and the resulting output distribution was sampled, with the interval corresponding to a $25$–$75$\% penetration probability displayed as error bars. This workflow is discussed in more detail in Ref. \cite{bursch_2026}. The second source is the input parameters and their temporal variation, which are found to be smaller than the uncertainties from the scaling law.

The blue curve (round markers) shows the experimental overlap metric calculated with GPEC for the imposed 3D fields, with error bars corresponding to 10\% of each value. This uncertainty was estimated as an upper bound using a Monte Carlo approach applied to three randomly selected discharges from this work. The magnetic data entering the equilibrium reconstruction were varied within their experimental uncertainties, and the resulting spread in the GPEC predictions was tracked.

The two curves intersect between the 2 and 3~kA cases, so the combined workflow, with GPEC providing the imposed  and Eq.~\ref{equ:Bursch_scaling} providing the threshold , correctly predicts the experimentally observed mode onset across the scan. This agreement lends credibility to applying the empirical multi-machine threshold scalings together with the more sophisticated stability modeling to predict mode onset in the ramp-up phase.

Moreover, the imposed $\delta$ increases linearly with the 3D coil current across the scan. Since GPEC computes the linear response of a given 2D equilibrium, $\delta$ is proportional to the coil current for a fixed equilibrium. Each point in Fig.~\ref{fig:subset_scan_modeling}, however, is computed from its own reconstructed equilibrium, and for the $3$~kA case this equilibrium was sampled after mode onset. A significant modification of the background equilibrium by the mode, e.g. of the $q$-profile, would therefore appear as a deviation from the linear trend. The absence of such a deviation indicates that the mode onset did not significantly modify the axisymmetric background equilibrium, which supports the use of 2D equilibria as inputs to the GPEC workflow even after mode onset.

\section{Database Approach}
\label{sec:database}
To strengthen these insights, the range of studied discharges was extended, scanning the discharges between discharge \#170,000 and \#200,000. While many discharges feature strong 3D coil currents during the plasma current ramp-up, these coils are typically operated in configurations intended for error field correction and, therefore, not usable in this work. Instead, this work requires discharges in which the 3D coils increase rather than compensate the error field, either unintentionally or in dedicated experiments such as the scan presented in Section~\ref{sec:analysis}. As this is undesirable for routine operation, such cases are rare. To this end, the selected dataset excludes discharges with typical EFC phases or vacuum discharges. The data mining then yields a small dataset of 12 discharges, listed in the appendix in Tab.~\ref{tab:studied_database}. All data points were sampled during the current ramp-up, with 8 of the 12 also featuring ramping 3D currents. For consistency, the sampling time was chosen $20$~ms before the start of the flat-top for cases without MHD activity, representative of the highest applied 3D currents, and $20$~ms before mode seeding for cases with MHD activity. All discharges considered here are dominated by imposed $n=1$ 3D fields. Discharges with dominant $n=3$ perturbations during the initial breakdown phase of the current ramp-up, such as those studied by Yang \textit{et al.}~\cite{yang_permanent_elm_tokamak}, were not found in a similar analysis to be susceptible to MHD mode seeding within the range of studied 3D coil currents.

The plasma parameters spanned by these data points are summarized in Tab.~\ref{tab:parameter_stats}. The dominant variations are in the density $n_e$ and the corresponding normalized pressure $\beta_n$, with minor variation in $B_t$ and $\kappa$. Meaningful variation in $R_0$ and $a$ would require a multi-machine study, beyond the scope of this work. Due to the transient ramp-up, the internal inductance, a measure of current peaking, varies by about 50\% and is relatively low compared with flat-top values. Rotation measurements were not available for all discharges due to the absence of the required diagnostic beam. Similarly, MSE measurements were unavailable for some discharges, as shown in Tab.~\ref{tab:studied_database}. In those cases, the equilibria used in the GPEC workflow were reconstructed with EFIT using magnetic measurements only, while MSE measurements were included as additional constraints on the current profile wherever available. In the latter cases, the MSE constraint changed $q_0$ only slightly.

\begin{figure}
    \centering
    \includegraphics[width=0.85\linewidth]{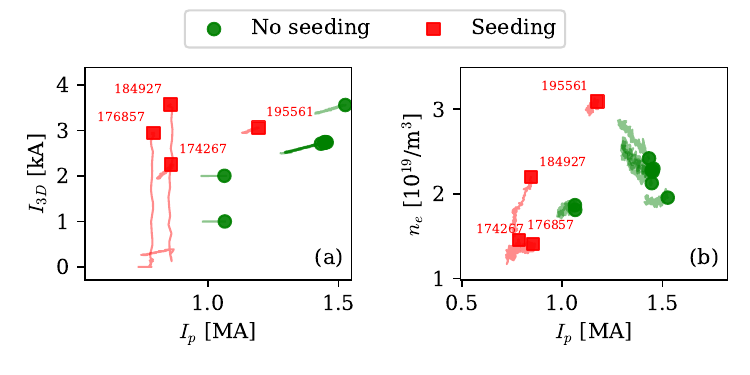}
    \caption{Overview of the explored parameter space. Each discharge is shown as a time series of samples through the ramp-up (a ``tail'' tracing its trajectory over $100$~ms), with the marker denoting the sampling time and the color and marker style indicating the experimental observation of mode seeding. Panel~(a) displays the plasma current $I_p$ versus 3D coil current $I_{3D}$. Panel~(b) shows $I_p$ versus line-averaged density $n_e$.}
    \label{fig:Ip_vs_ne}
\end{figure}

The sampling yields a range of combinations of $I_p$, $I_{3D}$, and $n_e$, illustrated in Fig.~\ref{fig:Ip_vs_ne}. Rather than a single point per discharge, each shot is drawn as a time series of slices, making its evolution through the ramp-up visible. The color coding and marker style indicate the experimentally observed mode seeding, classified with the workflow detailed in the previous section, involving compensation for the imposed 3D fields and subtraction of unperturbed reference discharges. Discharges with mode amplitudes exceeding the noise level during the perturbation and ramp-up were classified as seeded.

The trajectories in Panel~(a) suggest a trend among the seeded cases: two exhibit a jump in $I_{3D}$ at nearly constant $I_p$, and one a strong rise in $I_{3D}$ during the $I_p$ ramp. The remaining seeded case, however, shows a concurrent ramp of $I_p$ and $I_{3D}$ similar to two of the cases without seeding. This could indicate that the seeding is mainly controlled by the applied 3D coil current rather than by the plasma current or its ramp rate, consistent with a scenario in which the plasma, and correspondingly the threshold $\delta$, evolves slowly in the background while the imposed $\delta$ rises to exceed it. Cases with similar trajectories can, however, still differ in other parameters, which are accounted for when comparing the imposed and threshold $\delta$, and the studied dataset is too small and not homogeneous enough to draw firm conclusions from the trajectories alone.

Panel~(b) shows the $n_e$ for the same data points. All trajectories ramp toward higher $I_p$, while some of the seeded cases trend toward higher density and the cases without seeding toward lower density. This is instructive when read through the threshold scaling of Eq.~\ref{equ:Bursch_scaling}, where the threshold varies as $I_p^{-1.01}$ and $n_e^{0.56}$: along every trajectory the rising $I_p$ lowers the penetration threshold, whereas the rising density of the seeded cases would, on its own, raise it. That these discharges seed despite a density evolution working against penetration, while the cases without seeding remain stable even as their threshold falls with $I_p$, indicates that the imposed $I_{3D}$ drive dominates over the competing background-parameter trends. These insights are, however, limited by the relatively small number of discharges.

\begin{figure}
    \centering
    \includegraphics[width=0.6\linewidth]{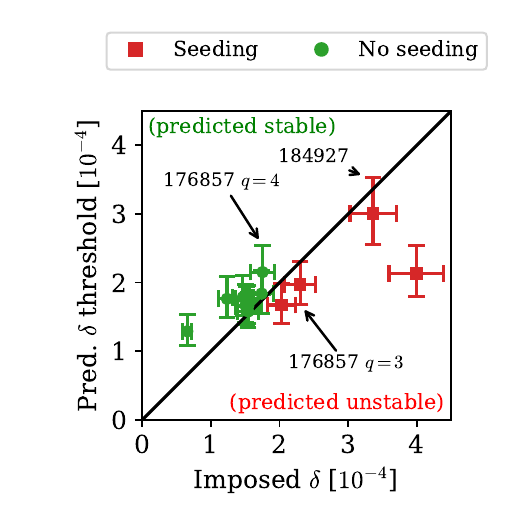}
    \caption{Comparison of the predicted and experimentally calculated overlap metric $\delta$. The solid line marks unity; points to the left (right) are predicted stable (unstable). Color coding indicates observed mode seeding.}
    \label{fig:database_GPEC}
\end{figure}

The overlap metric predicted by the error field penetration threshold scalings (y-axis) is compared with the imposed values from GPEC calculations (x-axis), see Fig.~\ref{fig:database_GPEC}. The solid 1:1 unity line separates predicted stable cases (left) from predicted unstable ones (right). Error bars on the y-axis are estimated with the Monte Carlo approach described above, and the x-axis uncertainties are taken as 10\%. The color coding and marker style indicate the experimentally observed mode seeding. In general, the workflow predicts the observed seeding correctly, with cases without seeding predominantly on the left and seeded cases on the right.\par

Moreover, three of the four seeded cases have an on-axis safety factor of $q_0\approx3$ at the sampling time, whereas the cases without seeding span a mixture of $q_0\approx3$ and $q_0\approx2$ (Tab.~\ref{tab:studied_database}). This indicates that the presence of the $q=3$ surface is necessary but not sufficient for seeding in these particular cases on DIII-D: consistent with the detailed analysis of \#176857, the $m/n=3/1$ surface is the dominant coupling channel, but seeding additionally requires the imposed $I_{3D}$ drive to exceed the threshold, as shown by the trajectories in Fig.~\ref{fig:Ip_vs_ne}.\par

Overall, this demonstrates that the workflow yields reliable predictions even under transient background conditions, evolving 3D coil currents, and the variations documented in Tab.~\ref{tab:parameter_stats}. Notably, without including the in-situ error fields in the calculations, only about half of the discharges were correctly classified, whereas including them via SURFMN recovers the correct classification across the database. This underlines the importance of accounting for in-situ error fields when assessing the penetration of additional, externally imposed 3D perturbations.

Multiple data points deserve comment. Two belong to discharge \#176857, the high-current case from the scan in Section~\ref{sec:analysis}, sampled as the $q=4$ and $q=3$ surfaces enter the core. At $q=4$, mode seeding is predicted to be marginal; at $q=3$, seeding is correctly predicted. This is consistent with insights from Fig.~\ref{fig:higher_n_numbers}(d).

\begin{figure}
    \centering
    \includegraphics[width=0.7\linewidth]{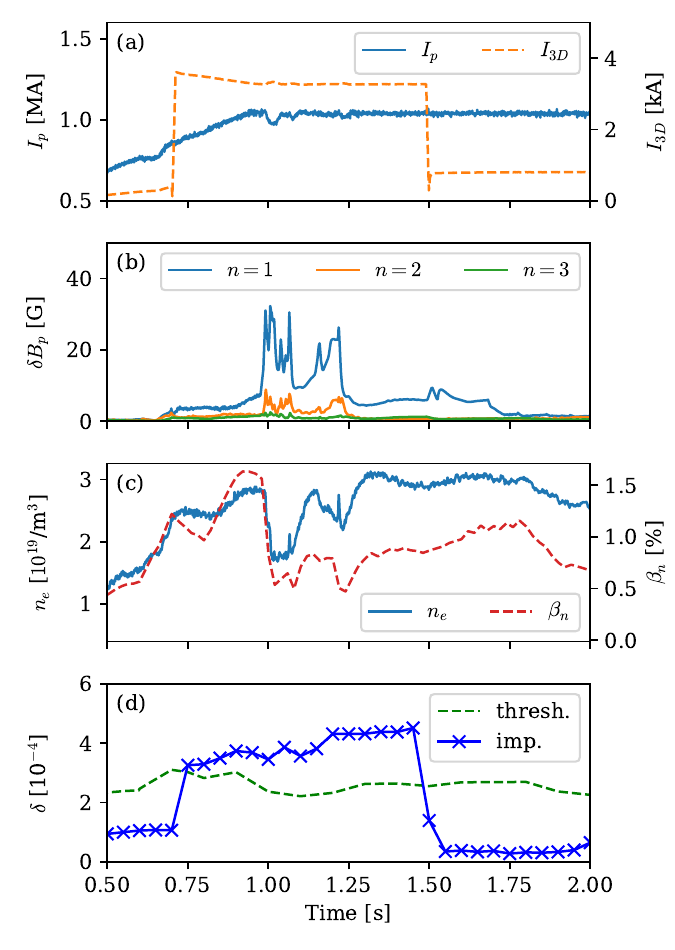}
    \caption{Time traces for discharge \#184927. Panel (a) shows the plasma current $I_p$ in blue and the applied C-coil current in orange with a dashed line. Panel (b) depicts the toroidal mode number decomposition of $\delta B_\theta$, showing transient $n=1$ activity correlated with the application of 3D fields and vanishing later in the discharge, indicative of the marginal seeding predicted in Fig.~\ref{fig:database_GPEC}. Panel (c) shows the measured density and $\beta_n$ over time. A time-dependent calculation of the predicted and imposed $\delta$ is shown in Panel (d).}
    \label{fig:higher_n_numbers_184927}
\end{figure}

In addition, there is a marker for discharge \#184927, which is found to be marginally seeded, see Fig.~\ref{fig:higher_n_numbers_184927} for details. Panel~(a) shows the plasma current (solid blue) and the applied C-coil current (dashed orange). Here the C-coils supply the applied perturbation, in contrast to their error-correction role in Section~\ref{sec:analysis}. The perturbation switches on during the ramp-up at about $0.75$~s and stays on until $\sim\!1.5$~s in the flat-top. As seen in Panel~(b), the perturbation immediately induces a measurable $n=1$ component in $\delta B_\theta$, which grows further once the ramp-up concludes near $1$~s and shoots up as the plasma current exhibits transients. Panel~(c) shows that this increase coincides with a strong drop in $n_e$ and $\beta_n$ around $1$~s, found to be a loss of ELMy H-mode as the heating decreases while strong gas puffing is applied (not shown here for brevity). As $n_e$ and $\beta_n$ recover from about $1.25$~s, the mode amplitude decreases again. After the C-coils switch off at $\sim\!1.5$~s, the mode seeding falls to the noise level by about $1.75$~s. Panel~(d) shows the time-dependent penetration threshold (dashed green) and the imposed 3D field (blue, with markers), which only marginally exceeds the threshold when the perturbation is switched on and then grows to exceed it by nearly a factor of two. As $n_e$ drops around $1$~s, the threshold falls too, widening the gap between imposed and threshold $\delta$ and coinciding with the increased mode seeding. Later the plasma stabilizes, $n_e$ and $\beta_n$ increase, and the plasma returns to marginal behavior, with the mode signature disappearing. While the precise mechanism behind this healing is not immediately clear, this case provides another example of a marginally seeded plasma recovering once the perturbation vanishes and the background parameters change.

\section{Summary}
\label{sec:summary}

This study investigates the impact of externally imposed non-axisymmetric error fields during the plasma current ramp-up time phase in the DIII-D tokamak, with particular relevance to future devices employing passive runaway electron mitigation coils.\par 

Detailed analysis of selected discharges demonstrates that sufficiently strong imposed $n=1$ perturbations can trigger error field penetration and seed MHD modes under transient conditions. In a dedicated scan, an apparent empirical threshold between $2$ and $3$~kA in applied 3D coil current is identified, above which mode seeding occurs. The imposed 3D fields initially couple to the higher-$q$ surfaces present early in the ramp and evolve toward lower rational surfaces as the discharge proceeds, showing some response near $q=4$, developing strongly once $q=3$ enters the plasma, and appearing predominantly as an $m/n=3/1$ mode by the end of the perturbation. These results show that the present error field penetration threshold scalings predict the onset of mode seeding, while the accompanying GPEC modeling of the experimentally imposed perturbation captures its coupling to the $q=3$ surface.\par

Extending this approach to a small database of 12 ramp-up discharges confirms these results across a range of plasma currents, densities, and applied 3D fields, suggesting that quasi-steady workflows can successfully predict EF thresholds during $I_p$ ramp-up. Displaying the discharge trajectories suggests that the seeding tends to be controlled by the applied 3D coil current rather than by the plasma current or its ramp rate. Consistently including the in-situ error fields of the device via SURFMN is found to be important: without them, only about half of the discharges are correctly classified. Notably, although the multi-machine penetration threshold scalings are intended to capture broad trends rather than a precise boundary, and carry substantial uncertainties, they predict the observed mode seeding in these ramp-up discharges with unexpectedly good accuracy.\par

In general, these results support the use of the applied workflow in the design and operation of future tokamaks, where transiently induced error fields from passive, conducting structures may be unavoidable. While the threshold scaling of Eq.~\ref{equ:Bursch_scaling} is derived from a multi-machine database, the present validation under ramp-up conditions is limited to DIII-D. Future work should therefore focus on extending the analysis to a wider parameter space and to other devices. Another important and fruitful direction is to study the effect of transient applied 3D fields in the plasma flat-top, by mining existing experiments, executing additional experiments where gaps are found, and supporting more detailed experiment-model validation efforts.

\appendix

\section{Appendix}

\begin{table}[h!]
\centering
\begin{tabular}{|c|c|c|c|c|}
\hline
Discharge \# & used & Time & Seeding & $q_0$ \\
\hline
176854 & I-coils & 0.900 & no & 2.6 (MSE)\\
176856 & I-coils & 0.900 & no & 2.1 (MSE)\\
176857 & I-coils & 0.599 & no & 3.3 (MSE)\\
176857 & I-coils & 0.750 & yes & 2.8 (MSE)\\
184927 & C-coils & 0.692 & yes & 2.1 \\
174267 & I-coils & 0.970 & yes & 2.2 \\
172347 & C-coils & 1.182 & no & 1.4 \\
172348 & C-coils & 1.190 & no & 1.4 \\
172349 & C-coils & 1.190 & no & 1.4 \\
172350 & C-coils & 1.190 & no & 1.4 \\
172351 & C-coils & 1.190 & no & 1.2 \\
176951 & I-coils & 1.350 & no & 1.3 (MSE)\\
195561 & C-coils & 1.700 & yes & 1.1 (MSE)\\
\hline
\end{tabular}
\caption{Studied database from Section \ref{sec:database}. Shown are the discharge number, the 3D coil type used to impose the perturbation, the sampled time point during the perturbation, and whether MHD activity was observed in the magnetic analysis. Values of $q_0$ obtained without an MSE constraint are subject to uncertainty and cannot be regarded as measurements. They are nonetheless listed here because they are the values that result from the equilibrium reconstruction and hence the stability calculations.}
\label{tab:studied_database}
\end{table}

\begin{table}[h!]
\centering
\begin{tabular}{|r|c|c|c|}
\hline
\textbf{} & \textbf{Min.} & \textbf{Avg.} & \textbf{Max.} \\
\hline
$n_e$ [1/m$^{3}$]     & 0.98 & 2.06 & 3.05 \\
$B_t$ [T]             & 1.68 & 1.98 & 2.15 \\
$R_0$ [m]             & 1.67 & 1.71 & 1.75 \\
$a$ [m]               & 0.56 & 0.58 & 0.59 \\
$\beta_n$             & 0.14 & 0.48 & 1.16 \\
$l_i$                 & 0.73 & 0.84 & 1.06 \\
$\kappa$              & 1.70 & 1.86 & 1.97 \\
$I_p$ [MA]            & 0.69 & 1.20 & 1.52 \\
$I_{3D}$ [kA]         & 0.98 & 2.48 & 3.22 \\
$\dot{I}_p$ [MA/s]    & 0.43 & 0.90 & 1.06 \\
$\dot{I}_{3D}$ [kA/s] & 0.00 & 1.42 & 2.34 \\
\hline
\end{tabular}
\caption{Minimum, average, and maximum values of selected parameters for the analyzed discharge database.}
\label{tab:parameter_stats}
\end{table}

\newpage

\printbibliography

\section*{Acknowledgments}
The authors would like to acknowledge the support by G. DeGrandchamp for the MSE analysis and N. Richner for supporting the magnetics analysis. This material is based upon work supported by the U.S. Department of Energy, Office of Science, Office of Fusion Energy Sciences, using the DIII-D National Fusion Facility, a DOE Office of Science user facility, under Award(s) DE-FC02-04ER54698, DE-FG02-04ER54761 and DE-SC0022270. This report was prepared as an account of work sponsored by an agency of the United States Government. Neither the United States Government nor any agency thereof, nor any of their employees, makes any warranty, express or implied, or assumes any legal liability or responsibility for the accuracy, completeness, or usefulness of any information, apparatus, product, or process disclosed, or represents that its use would not infringe privately owned rights. Reference herein to any specific commercial product, process, or service by trade name, trademark, manufacturer, or otherwise does not necessarily constitute or imply its endorsement, recommendation, or favoring by the United States Government or any agency thereof. The views and opinions of authors expressed herein do not necessarily state or reflect those of the United States Government or any agency thereof. Large Language Models were used during the preparation of this manuscript to assist with grammar checking and to improve readability. All scientific content, interpretations, and conclusions have been carefully verified by the authors.

\end{document}